\documentclass[sigconf,nonacm]{acmart}
\usepackage{array}
\usepackage{tikz}
\usetikzlibrary{arrows.meta,backgrounds,calc,fit,positioning}

\title[From Research Questions to Columns]
{From Research Questions to Columns: Operationalization-Aware Data Discovery}

\author{Houming Chen}
\affiliation{%
  \institution{University of Michigan}
  \city{Ann Arbor}
  \state{Michigan}
  \country{USA}
}
\email{houmingc@umich.edu}

\author{H. V. Jagadish}
\affiliation{%
  \institution{University of Michigan}
  \city{Ann Arbor}
  \state{Michigan}
  \country{USA}
}
\email{jag@umich.edu}

\begin{abstract}
Researchers often approach a data repository with an abstract concept and ask
which columns can measure it. Useful columns may not resemble the query; they
may matter only as complementary indicators in a defensible measure. This need
differs from schema linking and column retrieval, which begin from more explicit
needs and reward direct relevance. We define operationalization-aware data
discovery (OADD): given a broad question and a database, optionally under a
scope constraint, OADD jointly determines how focal concepts can be measured
with available data and identifies supporting columns.

Developing OADD methods requires examples for design and evaluation, but asking
researchers to supply conceptual questions and their columns is impractical. We
construct OADD-Bench by treating empirical papers as records of schema in use.
A question miner extracts and reframes a paper-supported question; a
paper-conditioned column miner reconstructs its measurements and grounds them
to database identifiers. We admit only mappings supported by the publication
and database documentation. OADD-Bench contains 160 questions from 111 papers
and 4,682 question--column labels. Each target records a measurement used in
published research; the paper supplies the precedent, while the miners extract
and ground it.

We evaluate lexical and neural retrieval, adapted schema-linking systems, and
large language model (LLM) OADD agents. Each method receives only a question,
permitted years, and dataset metadata; source papers are used only to construct
and document benchmark labels. At the largest output limit, direct retrieval
reaches at most 0.185 recall. The strongest schema-linking adaptation reaches
0.401 but remains optimized for a different objective; an OADD-directed agent
performs best at 0.465. Even this agent covers less than half the target columns,
showing that OADD remains an open problem.
\end{abstract}


\begin{document}
\maketitle

\section{Introduction}

A database organizes concrete facts, while a researcher approaching a data source has conceptual questions. Having the relevant observations in a database does
not make them discoverable from a scientific idea. Most column-search methods
look for a direct lexical or semantic relation between a query and a field. They do not consider the concretization of an abstract concept, the triangulation to it by means of related other variables, or other such adjustments needed to ``operationalize'' the concept of interest to the researcher.
As such, there is an \emph{operationalization gap}: deciding which
concrete observations can represent an abstract concept, whether or not any of them directly resemble the target concept syntactically or semantically.

\begin{figure}[!b]
  \centering
  \Description{Direct search finds no social-isolation field in HRS. An
  indirect path connects the concept to available columns through
  household size, proximity to children, religious attendance, and
  volunteering.}
  \resizebox{0.98\columnwidth}{!}{\begin{tikzpicture}[
  font=\sffamily\small,
  >={Latex[length=2.4mm]},
  box/.style={
    draw=black!55,
    rounded corners=4pt,
    line width=0.8pt,
    align=center,
    inner xsep=7pt,
    inner ysep=5pt
  },
  concept/.style={
    box,
    fill=blue!9,
    draw=blue!55!black,
    minimum width=0.40\columnwidth,
    minimum height=0.96cm
  },
  database/.style={
    box,
    fill=black!5,
    minimum width=0.29\columnwidth,
    minimum height=1.02cm
  },
  indicators/.style={
    box,
    fill=green!7,
    draw=green!45!black,
    minimum width=0.61\columnwidth,
    minimum height=1.62cm
  },
  chip/.style={
    draw=black!25,
    rounded corners=5pt,
    fill=white,
    font=\sffamily\scriptsize,
    inner xsep=4pt,
    inner ysep=2pt
  },
  edgelabel/.style={
    fill=white,
    rounded corners=2pt,
    inner xsep=4pt,
    inner ysep=2pt,
    align=center,
    font=\sffamily\scriptsize\bfseries
  }
]
  \node[concept] (concept) at (-2.25cm,0) {%
    {\scriptsize\color{blue!55!black}\textbf{CONCEPT OF INTEREST}}\\[2pt]
    {\normalsize\textbf{Social isolation}}
  };

  \node[database] (database) at (2.65cm,0) {%
      {\scriptsize\color{black!55}\textbf{DATABASE}}\\[3pt]
      {\normalsize\textbf{HRS}}
  };

  \node[indicators] (indicators) at (0.12cm,-2.00cm) {};

  \node[
    anchor=north,
    font=\sffamily\scriptsize\bfseries,
    text=green!35!black
  ] at ($(indicators.north)+(0,-0.15cm)$) {OBSERVABLE INDICATORS};

  \node[chip] at ($(indicators.center)+(-1.15cm,0.04cm)$)
    {household size};
  \node[chip] at ($(indicators.center)+(1.15cm,0.04cm)$)
    {proximity to children};
  \node[chip] at ($(indicators.center)+(-1.15cm,-0.45cm)$)
    {religious attendance};
  \node[chip] at ($(indicators.center)+(1.15cm,-0.45cm)$)
    {volunteering};

  \draw[->, line width=1.05pt, color=red!65!black]
    (concept.east) -- node[
      edgelabel,
      text=red!65!black,
      inner xsep=2pt,
      above=4pt,
      pos=0.54
    ] {direct search\\[-1pt]
       \normalfont\scriptsize\itshape no ready-made field}
    (database.west);

  \node[
    circle,
    draw=red!65!black,
    fill=white,
    text=red!65!black,
    font=\sffamily\bfseries,
    inner sep=1.5pt
  ] at ($(concept.east)!0.74!(database.west)$) {\(\times\)};

  \draw[->, line width=1.20pt, color=violet!65!black]
    (concept.south) -- node[
      edgelabel,
      text=violet!65!black,
      inner xsep=2pt,
      left=6pt,
      pos=0.48
    ] {can be measured by}
    ($(indicators.north west)!0.45!(indicators.north)$);

  \draw[->, line width=1.20pt, color=violet!65!black]
    ($(indicators.north)!0.55!(indicators.north east)$) -- node[
      edgelabel,
      text=violet!65!black,
      inner xsep=2pt,
      right=6pt,
      pos=0.50
    ] {exist in}
    (database.south);

\end{tikzpicture}}
  \caption{Operationalization connects a concept to available columns through
  observable indicators, even when direct search finds no field. Components
  follow Cenzer et al.~\cite{cenzer2025social}.}
  \label{fig:motivating-example}
\end{figure}
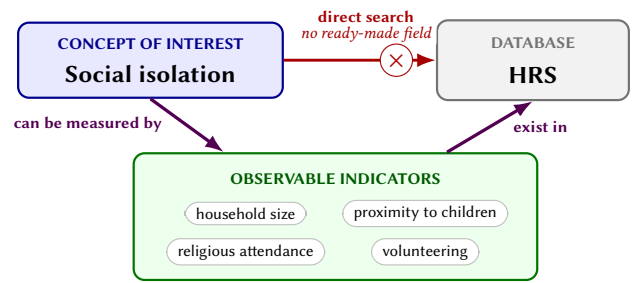

\noindent\textbf{Motivating example.} Consider the famous Health and Retirement Study (HRS), a longitudinal study of health, cognition, work, wealth, family, and social life~\cite{sonnega2014hrs}. Most HRS variables record recurring survey
responses, alongside many other types of data. Imagine a researcher investigating
whether social isolation is related to healthy aging. Searching for ``social
isolation'' finds no ready-made field. A fuzzy search may return variables
about Social Security; a semantic search may return a question about feeling
lonely. Neither reveals a defensible social-isolation measure. Yet Cenzer et
al.~\cite{cenzer2025social} construct one by combining marital status, household size, proximity to
children, religious attendance, and volunteering---all variables available in
HRS.
Individually, none of these variables states ``social isolation''; together, they
form a measure of it. Finding this bundle requires crossing the
operationalization gap, not merely improving lexical or semantic matching.
Figure~\ref{fig:motivating-example} summarizes this indirect mapping.

We call this task \emph{operationalization-aware data discovery} (OADD).
It is a common task for a substantial class of database users. Yet it has not been studied by the data management community. Its
closest existing problem is schema linking, commonly studied in
text-to-SQL~\cite{wang2020ratsql}. Both map a natural-language query to database
columns; they differ in what the query has already specified. A user asking
``How many flights went from Los Angeles to New York yesterday?'' may not know
the schema names, but has specified the event, endpoints, time, and desired
count. None of this is specified by a user seeking measures of social isolation.
An OADD system must therefore bridge the logical gap between a focal concept
and the observations that can measure it, using only data that
actually exist in the database. A scientifically sensible measure that
requires unavailable observations is not an answer. Neither is a list of
fields that fails to measure the focal concept.
Without this step, a usable database may appear not to contain the concept, or
a researcher may settle for a superficially related proxy; either failure can
change which studies appear feasible and how their concepts are measured.

In this paper, we identify the OADD problem and lay the groundwork for the data management community to address it.
OADD formalizes work that users perform manually when they
reconcile scientific concepts with available fields. It targets an early
point in the research workflow: broad question $\rightarrow$ focal-column candidates
$\rightarrow$ researcher validation and coding $\rightarrow$ analysis.
Section~\ref{sec:prob-defn} gives the formal problem definition.

Schema-linking methods typically assume that users know what data they need even if they do not know the exact schema names. They therefore seek direct relevance
and ignore fields valuable only within a measurement bundle.
Consequently, standard schema-linking methods tend to perform poorly on OADD
(see Section~\ref{sec:experiments}), motivating new methods for OADD.

Before we can develop new methods, though, we need to specify the OADD task more completely. What constitutes a good answer, and what should an OADD technique optimize? The desired answer is a set of columns that can operationalize the focal concept, often through relationships requiring domain expertise to recognize.
In short, we need a shared benchmark to make OADD concrete and give future
methods a common basis for comparison.

To create such a benchmark, we could directly ask human annotators to perform OADD:
interpret the idea, choose a defensible measure, and locate its exact columns.
This is expensive and requires multiple annotators with expertise in both the scientific domain and the database.

Our idea instead is to leverage records already produced through scientific practice. Peer-reviewed papers usually record
how domain researchers translated questions into measurements used in the reported
analyses. (This record may appear only in supplementary materials.)
Created by domain researchers for real studies and documented contemporaneously, these mappings are strong external positives rather than labels
invented for evaluation. Crucially, they are independent of any OADD system or
retrieval architecture. Our task then becomes to mine these implicit question-to-measurement
mappings and ground them in exact column identifiers. As we describe in
Section~\ref{sec:construction}, this is still much easier than recruiting large numbers of expert annotators.

Reconstructing the mapping from published papers is non-trivial because measurement details may be
scattered across the main text and supplementary materials, while exact
identifiers are often omitted. Construction must therefore synthesize paper
evidence while navigating documentation (for more than 122,000 fields in the case of HRS).

Recent large language model (LLM) agents make this construction strategy
practical: they can combine document understanding with multi-step reasoning
and tool use~\cite{openai2023gpt4,liu2024agentbench}. The published study
supplies the scientific precedent; the LLM finds its scattered measurement
description and connects it to HRS documentation. A \emph{question miner}
extracts a paper-supported question and reframes it for OADD; a \emph{column
miner} performs the more demanding measurement reconstruction and grounding.

Having constructed the benchmark, we use it (in Section~\ref{sec:experiments})
to ask whether familiar search and schema-linking
techniques already suffice. The OADD-directed agent achieves the highest
recall, but even its strongest tested model covers less than half of the targets.

To summarize, we make three main contributions:

\begin{itemize}
  \setlength{\itemsep}{1pt}
  \setlength{\parskip}{0pt}
  \setlength{\parsep}{0pt}
  \item \textbf{OADD.} We formulate a problem new to the data management
  community but familiar to researchers: jointly determining a scientific
  measurement and the available columns that can realize it.
  \item \textbf{OADD-Bench with publication-grounded construction.} We build and release OADD-Bench,\footnote{\url{https://github.com/umich-dbgroup/OADD-Bench}} a benchmark for OADD comprising 160
  publication-grounded questions with 4,682 labels from 111 papers in a
  122,324-column scientific database.
  Our evidence-constrained
  method grounds paper-reported measurements, preserves publication and
  identifier provenance, and abstains when a complete mapping is unsupported.
  \item \textbf{Diagnostic evaluation.} Target-relative recall across direct
  retrieval, five schema-linking adaptations, and one OADD-agent design with
  three model configurations exposes the gap between direct relevance and coverage
  of publication-grounded measurement bundles.
\end{itemize}

\begin{table*}[!t]
  \caption{Representative OADD question--column mappings. Permitted years are
  shown with each question; explanations make the intervening measurement
  decisions explicit, and target sets are shown in full.}
  \label{tab:benchmark-example}
  \centering
  \footnotesize
  \begin{tabular}{@{}p{0.18\textwidth}>{\raggedright\arraybackslash}p{0.21\textwidth}p{0.53\textwidth}@{}}
    \toprule
    Research question $q$ and scope constraint $C$ & Target columns $V$ &
    Explanation \\
    \midrule
    Could experiences of everyday discrimination be related to immune health
    in older adults?~\cite{kranz2026discrimination}
    {\footnotesize [permitted years: 2016]} &
    \texttt{PLB029A}--\texttt{PLB029F};\newline
    \texttt{PCD4T\_COUNT}, \texttt{PCD8T\_COUNT},
    \texttt{PBCELL\_COUNT};\newline
    \texttt{PCD4N\_COUNT}, \texttt{PCD4TEMRA\_COUNT},
    \texttt{PCD8N\_COUNT}, \texttt{PCD8TEMRA\_COUNT};\newline
    \texttt{PNAIVEB\_COUNT}, \texttt{PIGD\_PLUS\_MEMB\_COUNT},
    \texttt{PIGD\_MINUS\_MEMB\_COUNT} &
    \texttt{PLB029A}--\texttt{PLB029F} record frequencies of less respectful
    treatment, poorer service, being regarded as unintelligent or feared,
    threats or harassment, and worse medical treatment. Averaging them
    operationalizes everyday discrimination. The ten blood-assay fields count
    total, naive, and terminally differentiated T- and B-cell populations,
    operationalizing immune-cell composition. \\
    \addlinespace
    Could more positive views of aging protect everyday independence in later
    life?~\cite{su2025aging}
    {\footnotesize [permitted years: 2014, 2016]} &
    \texttt{OLB028B01}, \texttt{OLB028B02},
    \texttt{OLB028B3}--\texttt{OLB028B8};\newline
    \texttt{PG041}, \texttt{PG044}, \texttt{PG047}, \texttt{PG050},
    \texttt{PG059} &
    \texttt{OLB028B01}--\texttt{OLB028B8} ask about perceived decline,
    energy, usefulness, happiness, expectations, satisfaction with aging, and
    discontinued activities and disliked aspects of aging. Coding the
    responses in a common direction and averaging them operationalizes views
    of aging. The five \texttt{PG} fields record difficulty preparing meals,
    shopping, making phone calls, taking medications, and managing money;
    their sum is a 0--5 measure of impaired everyday independence. \\
    \bottomrule
  \end{tabular}
\end{table*}

\section{Operationalization-Aware Data Discovery}
\label{sec:oadd}

\subsection{Problem definition}
\label{sec:prob-defn}
Let $D$ be a database and $\mathrm{Col}(D)$ its uniquely identified columns.
A broad research question $q$ names one or more \emph{focal measurement
roles}---constructs or quantities examined as exposures, outcomes, mediators,
or moderators---without fully specifying how to observe them. An optional
scope constraint $C$ admits columns by time period, product, access tier, or
explicit subset, inducing
$\mathrm{Col}_{C}(D)=\{c\in\mathrm{Col}(D):C(c)\}$; without a restriction,
all columns are eligible.

An \emph{operationalization} $M$ maps each focal role $r$ to a nonempty set
$M(r)\subseteq\mathrm{Col}_{C}(D)$ whose fields, individually or jointly,
provide a scientifically defensible measure of $r$ in the context of $q$.
Its column set is $V(M)=\bigcup_r M(r)$.

\noindent\textbf{OADD problem.} Given $(q,D,C)$, find the column set $V(M)$
for a scientifically defensible operationalization $M$. A system
may expose or rank its interpretations, but the required result is a set of
column identifiers. Scientific defensibility is domain-dependent rather than
a schema property. Moreover, $M$ and $V(M)$ must be chosen jointly: an
unrealizable measure and a related-but-insufficient column list are both
invalid. This \emph{operationalization gap} is intrinsic rather than a
vocabulary mismatch: a useful column need not resemble $q$, and its role may
emerge only within a bundle.

OADD is the column-discovery stage of operationalization, not a complete study
design. It ends once the question has been grounded in a tractable set of fields
that a researcher can inspect. Coding, aggregation, modeling, and causal
identification remain downstream; covariates, weights, join keys, and
sample-flow rules are excluded unless the question makes them focal.

\subsection{Benchmark formulation}

OADD-Bench turns this task into a controlled empirical target.
Each OADD-Bench example is $b=(q,C,V)$, where $C$ specifies permitted
survey years and $V$ records one database-realizable operationalization of the focal
roles used in a published study. Documented use makes $V$ a strong positive
target. Other defensible measurements may exist, but the available database
constrains plausible alternatives: HRS social-isolation measures vary in
coverage but repeatedly use household ties, family contact, and social
participation~\cite{cenzer2025social,rosenberg2026endoflife}. The publication
establishes the scientific use and serves only as construction provenance.

At evaluation time, the publication is absent: the system addresses OADD
directly by selecting columns from the question and database.
$V$ supplies a known positive set for reproducible scoring: overlap shows that
the system found columns with demonstrated scientific use.

A benchmark target must be objectively checkable. Construct names or prose
plans would require semantic judgment at scoring time and leave the data
unresolved. We therefore ground $V$ in exact HRS identifiers, enabling direct
set comparison and field inspection. A reported recode or composite is traced
to source fields or withheld. The metadata covers 122,324 identifiers across
169 products and releases.

During prediction, evaluated systems receive the question, scope, output
budget, and HRS metadata; the paper-derived measurement plan, target columns,
and explanation are not method inputs. The released benchmark CSV provides
these targets and a concise operationalization explanation for scoring and
inspection. Companion JSON preserves detailed paper-use and
identifier-grounding records for audits without redistributing paper text.

\subsection{Recommended metrics}

Exact-column recall measures coverage of these demonstrated-use targets under
controlled output budgets, not exhaustive correctness over every possible
operationalization. For $R_q=|V_q|$, we report macro
Recall@$R$, Recall@$2R$, and Recall@$5R$, allowing at most $R_q$, $2R_q$, or
$5R_q$ returned columns. These are evaluation budgets, not a deployment
assumption that target size is known. They normalize different target sizes,
prevent return-all strategies, and let $5R$ accommodate broader candidate
sets. Methods may return fewer columns, and scoring uses their actual output.
Across many questions, low recall at $5R$ means repeatedly missing fields
already demonstrated to support the focal concepts.

\section{Constructing OADD-Bench}
\label{sec:construction}

\subsection{Construction overview}

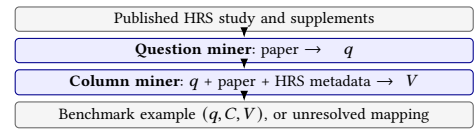
\begin{figure}[b]
\centering
\Description{A published HRS study enters a question miner that extracts a
scientific question. A paper-conditioned column miner then reconstructs the
measurement and grounds it to HRS identifiers, yielding a benchmark example
or an unresolved mapping.}
\begin{tikzpicture}[
  node distance=0.8mm,
  flow/.style={-{Latex[length=1.4mm]}, semithick},
  io/.style={draw=black!60, fill=black!4, rounded corners=1.5pt,
    align=center, font=\scriptsize, inner sep=1.8pt,
    text width=0.70\columnwidth},
  agent/.style={draw=blue!55!black, fill=blue!7, rounded corners=1.5pt,
    align=center, font=\scriptsize, inner sep=1.8pt,
    text width=0.70\columnwidth}
]
  \node[io] (paper) {Published HRS study and supplements};
  \node[agent, below=of paper] (qminer)
    {\textbf{Question miner}: paper $\rightarrow q$};
  \node[agent, below=of qminer] (cminer)
    {\textbf{Column miner}: $q$ + paper + HRS metadata $\rightarrow V$};
  \node[io, below=of cminer] (output)
    {Benchmark example $(q,C,V)$, or unresolved mapping};
  \draw[flow] (paper) -- (qminer);
  \draw[flow] (qminer) -- (cminer);
  \draw[flow] (cminer) -- (output);
\end{tikzpicture}
\caption{Papers provide construction evidence; evaluated systems later receive
only $q$, $C$, and the database metadata.}
\label{fig:pipeline}
\end{figure}

Published studies provide scientifically grounded labels, but their
measurement descriptions must be mapped to exact HRS identifiers. Rosenberg et
al., for example, describe social isolation as a 15-question composite and
place its components in a supplementary figure~\cite{rosenberg2026endoflife}.
One is ``Number of children within 10 miles,'' but neither source states that
this maps to \texttt{QE012} in 2018 and \texttt{KE012}--\texttt{PE012} in
earlier waves. Grounding it requires joining scattered paper evidence with HRS
documentation. Modern LLMs make this supervision extractable at scale through
scientific interpretation and iterative metadata
navigation~\cite{majumder2024discoverybench,yang2024llmmeasure}. We use a question miner
and a paper-conditioned column miner, retaining only evidence-complete mappings
(Figure~\ref{fig:pipeline}).

\subsection{Question miner}

The question miner extracts the questions investigated in a paper and
expresses each as a broad, early-stage question, preserving its focal concepts,
relation, population, and scope. A paper may contribute multiple questions.
This document-understanding task is related to scientific summarization and
research-question extraction~\cite{cachola2020tldr,taslimi2025rq}.

It also removes measurement details that would reveal the answer. For example,
the first question in Table~\ref{tab:benchmark-example} asks whether everyday
discrimination is related to immune health without naming the six
treatment-report fields or the T- and B-cell counts. Including those names
would reduce the task to direct schema linking. The relation and population
remain; the measurements have not yet been chosen.

\subsection{Column miner}

The column miner receives the question, the paper and its supplements, and official
HRS metadata. It uses a coarse-to-fine process to turn scattered narrative
evidence into a small evidence packet and then exact identifiers. All LLM
passes use GPT-5.5 with schema-constrained outputs.

\noindent\textbf{1. Reconstruct the measurement.} The miner identifies the
focal roles in the question and synthesizes how the paper observes each one,
including its components, scope, product clues, and supporting passages across
the main text and supplements. It excludes nonfocal covariates and cannot
propose identifiers at this stage.

\noindent\textbf{2. Retrieve candidate evidence.} Scale names, observable
indicators, and codebook wording may differ sharply, so the miner issues
separate queries for named measures, observable components, and likely
codebook phrasing. Lexical retrieval is combined across identifiers, labels,
question text, and product metadata; scope and repeated-wave structure recover
compatible fields and companion items. Per-measurement capacity prevents an
easy lexical match from displacing less obvious components.

\noindent\textbf{3. Build a reviewable evidence packet.} Retrieval is too
brittle to select labels, while presenting every field with full codebook
evidence is impractical. An LLM screening pass removes clearly unrelated
candidates but retains plausible raw items, released scores, routing fields,
companions, and repeated-wave variants. Full evidence is loaded only for this
shortlist.

\noindent\textbf{4. Ground or abstain.} A final pass jointly examines the
measurement record and shortlisted codebook evidence to resolve scope, units,
routing, and raw versus released fields. It may select only packet-supported
identifiers. If a focal component or derived-field lineage remains uncertain,
the mapping stays unresolved. Deterministic checks reject identifiers absent
from the evidence, non-HRS targets, inconsistent mappings, and duplicates.

\subsection{Evidence-based selection and scale}

Our publication frame is a June 2026 crawl of the public HRS bibliography,
yielding 9,954 heterogeneous records~\cite{hrsbibliography2026}. We pursued
full text through two routes: a systematic 500-record sample, screened for
likely analytic HRS use and public availability, and unambiguous open-access
Europe PMC records title-matched to the bibliography. Through public or
authorized routes, we attempted 861 distinct records and obtained 450
full-text documents usable for construction.

An HRS bibliography entry does not by itself yield an auditable label: some
papers do not report an empirical HRS analysis, while others omit the details
needed to reconstruct their focal measurements. We therefore select the
acquired papers and their candidate questions in two stages.

The first stage determines \emph{paper eligibility}. An accessible,
peer-reviewed article must report an empirical HRS analysis and describe an
analytically central measure and its HRS scope well enough to reconstruct its
observable components. Eligibility does not depend on miner success, metadata
coverage, or retrieval difficulty. Of 450 acquired full texts, 364 pass.

The second stage operates on candidate question--column examples; one paper
may support several. To focus this challenge set on cases where a bundle must
be discovered, an admitted example must ground at least two HRS columns. This
selection defines the benchmark, not OADD itself, which may have a
single-column answer. Each example must support every focal measurement in the
released question with publication evidence. When a paper studies several
relationships, a candidate question may focus on one supported relationship
rather than summarize the entire paper; it cannot introduce a construct or
guess an unresolved identifier or derived-variable lineage. Each selected
identifier must be supported by loaded HRS documentation.

The miners produce 448 candidate mappings; 140 remain after the two selection
stages and a final consistency screening. We additionally include 20 annotated
examples from distinct papers under the same evidence and grounding
requirements. Together they form an evidence-complete 160-example release,
not a sample of all HRS studies.

\noindent\textbf{Secondary consistency check.} For these 20 examples,
an annotator reviews and edits structured evidence prefills created
from the papers, supplements, and HRS documentation. We compare the resulting
annotations with the column-miner reconstructions across 640 labels. The miner covers 0.942 of
the annotated labels, while 0.946 of its labels are retained by the
annotations, yielding 0.933 macro set F1. Human annotations are not treated as
gold: this reconstruction itself requires scientific and HRS expertise.
Agreement therefore checks consistency between two reconstructions; the
publication supplies the scientific precedent.

The resulting benchmark contains 160 questions from 111 papers. Its 4,682
question--column label incidences cover 2,032 distinct HRS identifiers and 873
paper-documented measurement components. Target sets range from 2 to 136
columns (median 21; mean 29.3).

\section{Testing OADD Systems}
\label{sec:experiments}

\subsection{Evaluation design}

We compare direct retrieval, adapted schema linking, and LLM-based discovery.
Every system receives only the question, permitted years, an output
limit, and HRS metadata; paper identifiers, provenance, and measurement
descriptions are hidden. This fixes a reproducible experimental boundary
rather than restricting OADD generally. LLM conditions use metadata-search
tools but no paper or Web retrieval. The accompanying artifact includes the
benchmark CSV, detailed JSON audit records, raw HRS documentation, baselines,
and scoring code.

\begin{table}[t]
\caption{Question-only macro exact-column recall on all 160 questions.}
\label{tab:benchmark-retrieval}
\centering\footnotesize
\setlength{\tabcolsep}{3pt}
\begin{tabular}{@{}lrrr@{}}
\toprule
& \multicolumn{3}{c}{Macro exact-column recall}\\
\cmidrule(lr){2-4}
Method & Recall@$R$ & Recall@$2R$ & Recall@$5R$\\
\midrule
\multicolumn{4}{@{}l}{\emph{Direct column retrieval}}\\
BM25~\cite{robertson2009bm25} & 0.039 & 0.069 & 0.129\\
TF--IDF~\cite{salton1988term} & 0.038 & 0.071 & 0.127\\
BGE-base~\cite{xiao2023cpack} & 0.051 & 0.104 & 0.185\\
SPLADE++~\cite{formal2022spladepp} & 0.056 & 0.087 & 0.159\\
Four-way rank fusion~\cite{cormack2009rrf} & 0.068 & 0.111 & 0.179\\
\addlinespace[2pt]
\multicolumn{4}{@{}l}{\emph{Schema linking}}\\
RESDSQL ranker~\cite{li2023resdsql} & 0.012 & 0.021 & 0.054\\
CRUSH4SQL~\cite{kothyari2023crush4sql} & 0.020 & 0.028 & 0.048\\
Bidirectional~\cite{nahid2026rethinking} & 0.132 & 0.173 & 0.198\\
LinkAlign~\cite{wang2025linkalign} & 0.156 & 0.177 & 0.203\\
AutoLink~\cite{wang2026autolink} & 0.185 & 0.282 & 0.401\\
\addlinespace[2pt]
\multicolumn{4}{@{}l}{\emph{LLM agents}}\\
GPT-5.4 nano & 0.232 & 0.270 & 0.272\\
GPT-5.4 mini & 0.259 & 0.305 & 0.319\\
\textbf{GPT-5.5} & \textbf{0.399} & \textbf{0.458} & \textbf{0.465}\\
\bottomrule
\end{tabular}
\end{table}

\noindent\textbf{HRS representation.} Each searchable record contains its
identifier, label, codebook question text, and product metadata. Retrieval
groups repeated-wave equivalents and expands permitted-year columns up to the
output limit. Schema linkers receive the same metadata as 169 wide product or
release tables with 126,091 table--column occurrences. Both representations
resolve to official identifiers.

\noindent\textbf{Retrieval systems.} BM25 and TF--IDF test lexical matching;
SPLADE++ and BGE-base test learned sparse and dense retrieval. Four-way
reciprocal-rank fusion combines their rankings.

\noindent\textbf{Schema-linking systems.} We adapt five text-to-SQL methods.
The released Spider-trained RESDSQL schema-item classifier reranks
dense-prefiltered HRS packets without its SQL decoder~\cite{li2023resdsql}.
Four LLM adaptations use CRUSH4SQL's hypothetical-schema retrieval,
LinkAlign's gap-directed rewrite, AutoLink's iterative exploration, and
bidirectional table/column linking~\cite{kothyari2023crush4sql,
wang2025linkalign,wang2026autolink,nahid2026rethinking}. We retain these
linking mechanisms, map outputs to HRS identifiers, and omit SQL and value
tools. These are OADD adaptations, not native text-to-SQL reproductions.
CRUSH4SQL, LinkAlign, and AutoLink use GPT-5.5; bidirectional uses GPT-5.4 mini.

\noindent\textbf{OADD agents.} The agent iteratively formulates metadata
queries, inspects hybrid retrieval results, and returns exact identifiers. Its
prompt requests coherent measurement bundles without prescribing an
operationalization. We run this design with GPT-5.4 nano, GPT-5.4 mini, and
GPT-5.5. Independent $R$, $2R$, and $5R$ runs may return fewer than their
respective limits, so recall need not be monotone. Outputs are validated, deduplicated,
and charged to the same budget.

\subsection{Results and analysis}

\noindent\textbf{Direct similarity covers few benchmark targets.}
Across the lexical, dense, learned-sparse, and fused rankings in
Table~\ref{tab:benchmark-retrieval}, no direct retrieval method exceeds 0.185
at $5R$. Thus, even with an allowance five times the reference-set size,
direct question--field matching recovers less than one fifth of the documented
target columns on average. This is consistent with the operationalization gap:
useful fields need not directly resemble the question.

\noindent\textbf{Schema linking optimizes a different target.}
Schema linking is a closer comparison because it also maps language to schema
elements, but a text-to-SQL question typically already expresses the required
quantities. It therefore optimizes direct query--column relevance, whereas
OADD must determine which observations could measure a concept whose
operationalization remains unspecified. The adaptations vary widely: the
released RESDSQL ranker reaches 0.054 at $5R$, while the strongest LLM
adaptation, AutoLink, reaches 0.401, compared with 0.465 for the OADD agent,
whose performance we analyze in the following paragraph. A paired bootstrap
over source papers estimates this mean difference at 0.063 (95\% CI
[0.004, 0.121]). These scores measure how schema-linking mechanisms transfer
to OADD, not their native text-to-SQL performance.

\noindent\textbf{OADD-directed agents perform best, but remain incomplete.}
Higher column-level recall does not imply complete measurement bundles. At
$5R$, the GPT-5.5 agent retrieves at least one field for 61.0\% of a question's
measurement components on average and every field for 37.0\%. The corresponding
partial and complete rates are 59.0\% and 29.6\% for AutoLink, and 21.5\% and
9.8\% for BGE-base. The agent retrieves at least one field for every component
in only 50 of 160 questions (AutoLink: 43).

The remaining gap is not simply an output-budget effect. From $2R$ to $5R$,
the strongest agent's average output rises only from 33.3 to 39.9 columns,
against an average $5R$ allowance of 146.3, while recall moves from 0.458 to
0.465. Model choice also matters within the fixed agent design: Recall@$5R$ is
0.272 with GPT-5.4 nano, 0.319 with GPT-5.4 mini, and 0.465 with GPT-5.5. Yet
even the strongest tested system misses more than half of the documented
target columns.

\section{Related Work}

\textbf{Schema linking and data discovery.} Schema linking grounds
natural-language requirements in tables and columns, usually for
text-to-SQL~\cite{wang2020ratsql}. OADD shares the question-to-schema
interface, but begins before the required observations have been specified.
This logical gap is the object of the task: useful columns may be indirect
indicators whose role emerges only within a bundle, rather than columns that
directly resemble the query. RESDSQL, CRUSH4SQL, LinkAlign, AutoLink, and
bidirectional retrieval are therefore natural mechanisms to test, but their
benchmarks begin from more explicit requirements
~\cite{li2023resdsql,kothyari2023crush4sql,wang2025linkalign,wang2026autolink,
nahid2026rethinking}. Data-discovery systems retrieve tables and fields from
metadata, values, relations, and learned representations
~\cite{fernandez2018aurum,fan2023starmie}; recent LLM systems extend semantic
dataset and field search~\cite{viswanathan2023datafinder,ding2025pneuma,
green2026llmsearch}. Their usual item-wise relevance objectives do not test whether a
set of fields jointly realizes a scientific measurement.
SDRQuerier maps text to predefined variables in one harmonized survey database
and visualizes availability~\cite{tu2023sdrquerier}; it leaves measurement
assembly to users rather than discovering indirect bundles.

\textbf{Survey linking and scientific measurement.} Prior work links a named
phenomenon or publication passage to survey items
~\cite{dulisch2015query,tsereteli2022svident,tsereteli2024sil}. OADD instead
begins with a broad question, must infer the intervening measurement for its
focal roles, and grounds the resulting bundle to exact columns. LLMs also induce measures and derive
workflows~\cite{yang2024llmmeasure,majumder2024discoverybench}; these can supply
components, but OADD makes their joint realizability in the available database
part of the measurement choice.
Concurrent work catalogs operationalization failures in executable agent
workflows~\cite{mahmud2026semanticgap}; OADD instead asks which fields can
measure an unspecified scientific concept.

\section{Conclusion and Remarks}

We introduced OADD, the data-discovery problem of jointly reasoning about
scientific measurement and the database columns that can realize it.
OADD-Bench makes the problem testable with 160 publication-grounded questions
from 111 papers and 4,682 question--column labels. At $5R$, direct retrieval
reaches only 0.185 recall. The strongest schema-linking adaptation reaches
0.401 despite optimizing a different target; the OADD-directed agent reaches
0.465 but still covers less than half of the target columns. OADD-Bench provides a common basis for
developing stronger systems, including bundle-level discovery over repeated
waves, provenance-aware grounding, and interactive researcher validation.

\balance
\bibliographystyle{ACM-Reference-Format}
\bibliography{references}


\begin{thebibliography}{33}


\ifx \showCODEN    \undefined \def \showCODEN     #1{\unskip}     \fi
\ifx \showISBNx    \undefined \def \showISBNx     #1{\unskip}     \fi
\ifx \showISBNxiii \undefined \def \showISBNxiii  #1{\unskip}     \fi
\ifx \showISSN     \undefined \def \showISSN      #1{\unskip}     \fi
\ifx \showLCCN     \undefined \def \showLCCN      #1{\unskip}     \fi
\ifx \shownote     \undefined \def \shownote      #1{#1}          \fi
\ifx \showarticletitle \undefined \def \showarticletitle #1{#1}   \fi
\ifx \showURL      \undefined \def \showURL       {\relax}        \fi
\providecommand\bibfield[2]{#2}
\providecommand\bibinfo[2]{#2}
\providecommand\natexlab[1]{#1}
\providecommand\showeprint[2][]{arXiv:#2}

\bibitem[Balaka et~al\mbox{.}(2025)]%
        {ding2025pneuma}
\bibfield{author}{\bibinfo{person}{Muhammad Imam~Luthfi Balaka},
  \bibinfo{person}{David Alexander}, \bibinfo{person}{Qiming Wang},
  \bibinfo{person}{Yue Gong}, \bibinfo{person}{Adila Krisnadhi}, {and}
  \bibinfo{person}{Raul Castro~Fernandez}.} \bibinfo{year}{2025}\natexlab{}.
\newblock \showarticletitle{Pneuma: Leveraging {LLM}s for Tabular Data
  Representation and Retrieval in an End-to-End System}.
\newblock \bibinfo{journal}{\emph{Proceedings of the ACM on Management of
  Data}} \bibinfo{volume}{3}, \bibinfo{number}{3}, Article
  \bibinfo{articleno}{200} (\bibinfo{year}{2025}),
  \bibinfo{numpages}{28}~pages.
\newblock
\href{https://doi.org/10.1145/3725337}{doi:\nolinkurl{10.1145/3725337}}


\bibitem[Cachola et~al\mbox{.}(2020)]%
        {cachola2020tldr}
\bibfield{author}{\bibinfo{person}{Isabel Cachola}, \bibinfo{person}{Kyle Lo},
  \bibinfo{person}{Arman Cohan}, {and} \bibinfo{person}{Daniel Weld}.}
  \bibinfo{year}{2020}\natexlab{}.
\newblock \showarticletitle{{TLDR}: Extreme Summarization of Scientific
  Documents}. In \bibinfo{booktitle}{\emph{Findings of the Association for
  Computational Linguistics: EMNLP 2020}}. \bibinfo{publisher}{Association for
  Computational Linguistics}, \bibinfo{pages}{4766--4777}.
\newblock
\href{https://doi.org/10.18653/v1/2020.findings-emnlp.428}{doi:\nolinkurl{10.18653/v1/2020.findings-emnlp.428}}


\bibitem[Cenzer et~al\mbox{.}(2025)]%
        {cenzer2025social}
\bibfield{author}{\bibinfo{person}{Irena Cenzer},
  \bibinfo{person}{Jacqueline~M. Torres}, \bibinfo{person}{Yulin Yang},
  \bibinfo{person}{Karla~Renata Flores~Romero}, \bibinfo{person}{Mary~C.
  Thoma}, {and} \bibinfo{person}{Ashwin~A. Kotwal}.}
  \bibinfo{year}{2025}\natexlab{}.
\newblock \showarticletitle{Development and Evaluation of a Brief Social
  Isolation Measure in the Nationally-Representative Health and Retirement
  Study}.
\newblock \bibinfo{journal}{\emph{Journal of the American Geriatrics Society}}
  \bibinfo{volume}{73}, \bibinfo{number}{12} (\bibinfo{year}{2025}),
  \bibinfo{pages}{3807--3813}.
\newblock
\href{https://doi.org/10.1111/jgs.70056}{doi:\nolinkurl{10.1111/jgs.70056}}


\bibitem[Cormack et~al\mbox{.}(2009)]%
        {cormack2009rrf}
\bibfield{author}{\bibinfo{person}{Gordon~V. Cormack}, \bibinfo{person}{Charles
  L.~A. Clarke}, {and} \bibinfo{person}{Stefan Buettcher}.}
  \bibinfo{year}{2009}\natexlab{}.
\newblock \showarticletitle{Reciprocal Rank Fusion Outperforms Condorcet and
  Individual Rank Learning Methods}. In \bibinfo{booktitle}{\emph{Proceedings
  of the 32nd International ACM SIGIR Conference on Research and Development in
  Information Retrieval}}. \bibinfo{pages}{758--759}.
\newblock
\href{https://doi.org/10.1145/1571941.1572114}{doi:\nolinkurl{10.1145/1571941.1572114}}


\bibitem[Dulisch et~al\mbox{.}(2015)]%
        {dulisch2015query}
\bibfield{author}{\bibinfo{person}{Nadine Dulisch},
  \bibinfo{person}{Andreas~Oskar Kempf}, {and} \bibinfo{person}{Philipp
  Schaer}.} \bibinfo{year}{2015}\natexlab{}.
\newblock \showarticletitle{Query Expansion for Survey Question Retrieval in
  the Social Sciences}. In \bibinfo{booktitle}{\emph{Research and Advanced
  Technology for Digital Libraries: TPDL 2015}}. \bibinfo{publisher}{Springer},
  \bibinfo{pages}{28--39}.
\newblock
\href{https://doi.org/10.1007/978-3-319-24592-8_3}{doi:\nolinkurl{10.1007/978-3-319-24592-8_3}}


\bibitem[Fan et~al\mbox{.}(2023)]%
        {fan2023starmie}
\bibfield{author}{\bibinfo{person}{Grace Fan}, \bibinfo{person}{Jin Wang},
  \bibinfo{person}{Yuliang Li}, \bibinfo{person}{Dan Zhang}, {and}
  \bibinfo{person}{Ren{\'e}e~J. Miller}.} \bibinfo{year}{2023}\natexlab{}.
\newblock \showarticletitle{Semantics-aware Dataset Discovery from Data Lakes
  with Contextualized Column-based Representation Learning}.
\newblock \bibinfo{journal}{\emph{Proceedings of the VLDB Endowment}}
  \bibinfo{volume}{16}, \bibinfo{number}{7} (\bibinfo{year}{2023}).
\newblock


\bibitem[Fernandez et~al\mbox{.}(2018)]%
        {fernandez2018aurum}
\bibfield{author}{\bibinfo{person}{Raul~Castro Fernandez},
  \bibinfo{person}{Ziawasch Abedjan}, \bibinfo{person}{Samuel Madden}, {and}
  \bibinfo{person}{Michael Stonebraker}.} \bibinfo{year}{2018}\natexlab{}.
\newblock \showarticletitle{Aurum: A Data Discovery System}. In
  \bibinfo{booktitle}{\emph{2018 IEEE 34th International Conference on Data
  Engineering}}. \bibinfo{pages}{1001--1012}.
\newblock
\href{https://doi.org/10.1109/ICDE.2018.00094}{doi:\nolinkurl{10.1109/ICDE.2018.00094}}


\bibitem[Formal et~al\mbox{.}(2022)]%
        {formal2022spladepp}
\bibfield{author}{\bibinfo{person}{Thibault Formal}, \bibinfo{person}{Carlos
  Lassance}, \bibinfo{person}{Benjamin Piwowarski}, {and}
  \bibinfo{person}{St{\'e}phane Clinchant}.} \bibinfo{year}{2022}\natexlab{}.
\newblock \showarticletitle{From Distillation to Hard Negative Sampling: Making
  Sparse Neural {IR} Models More Effective}. In
  \bibinfo{booktitle}{\emph{Proceedings of the 45th International ACM SIGIR
  Conference on Research and Development in Information Retrieval}}.
  \bibinfo{publisher}{Association for Computing Machinery},
  \bibinfo{pages}{2353--2359}.
\newblock
\href{https://doi.org/10.1145/3477495.3531857}{doi:\nolinkurl{10.1145/3477495.3531857}}


\bibitem[Green et~al\mbox{.}(2026)]%
        {green2026llmsearch}
\bibfield{author}{\bibinfo{person}{Mark Green}, \bibinfo{person}{Maura
  Halstead}, \bibinfo{person}{Caroline Jay}, \bibinfo{person}{Richard
  Kingston}, \bibinfo{person}{Alex Singleton}, {and} \bibinfo{person}{David
  Topping}.} \bibinfo{year}{2026}\natexlab{}.
\newblock \showarticletitle{Comparing How Large Language Models Perform against
  Keyword-Based Searches for Social Science Research Data Discovery}.
\newblock \bibinfo{journal}{\emph{arXiv preprint arXiv:2601.19559}}
  (\bibinfo{year}{2026}).
\newblock
\href{https://doi.org/10.48550/arXiv.2601.19559}{doi:\nolinkurl{10.48550/arXiv.2601.19559}}


\bibitem[{Health and Retirement Study}(2026)]%
        {hrsbibliography2026}
\bibfield{author}{\bibinfo{person}{{Health and Retirement Study}}.}
  \bibinfo{year}{2026}\natexlab{}.
\newblock \bibinfo{title}{{HRS} Bibliography}.
\newblock
\shownote{Accessed June 2026}.
\newblock
\urldef\tempurl%
\url{https://hrs.isr.umich.edu/publications/biblio/}
\showURL{%
\tempurl}


\bibitem[Kothyari et~al\mbox{.}(2023)]%
        {kothyari2023crush4sql}
\bibfield{author}{\bibinfo{person}{Mayank Kothyari}, \bibinfo{person}{Dhruva
  Dhingra}, \bibinfo{person}{Sunita Sarawagi}, {and} \bibinfo{person}{Soumen
  Chakrabarti}.} \bibinfo{year}{2023}\natexlab{}.
\newblock \showarticletitle{{CRUSH4SQL}: Collective Retrieval Using Schema
  Hallucination for {Text2SQL}}. In \bibinfo{booktitle}{\emph{Proceedings of
  the 2023 Conference on Empirical Methods in Natural Language Processing}}.
  \bibinfo{pages}{14054--14066}.
\newblock
\href{https://doi.org/10.18653/v1/2023.emnlp-main.868}{doi:\nolinkurl{10.18653/v1/2023.emnlp-main.868}}


\bibitem[Kranz et~al\mbox{.}(2026)]%
        {kranz2026discrimination}
\bibfield{author}{\bibinfo{person}{Emiko~O. Kranz}, \bibinfo{person}{Jemar~R.
  Bather}, \bibinfo{person}{Xiaoyan Zhang}, \bibinfo{person}{Virginia~W.
  Chang}, \bibinfo{person}{Steven~W. Cole}, {and} \bibinfo{person}{Adolfo~G.
  Cuevas}.} \bibinfo{year}{2026}\natexlab{}.
\newblock \showarticletitle{Discrimination Exposure and Lymphocyte
  Differentiation: Results from the Health and Retirement Study}.
\newblock \bibinfo{journal}{\emph{Brain, Behavior, \& Immunity - Health}}
  \bibinfo{volume}{52} (\bibinfo{year}{2026}), \bibinfo{pages}{101170}.
\newblock
\href{https://doi.org/10.1016/j.bbih.2026.101170}{doi:\nolinkurl{10.1016/j.bbih.2026.101170}}


\bibitem[Li et~al\mbox{.}(2023)]%
        {li2023resdsql}
\bibfield{author}{\bibinfo{person}{Haoyang Li}, \bibinfo{person}{Jing Zhang},
  \bibinfo{person}{Cuiping Li}, {and} \bibinfo{person}{Hong Chen}.}
  \bibinfo{year}{2023}\natexlab{}.
\newblock \showarticletitle{{RESDSQL}: Decoupling Schema Linking and Skeleton
  Parsing for Text-to-{SQL}}. In \bibinfo{booktitle}{\emph{Proceedings of the
  AAAI Conference on Artificial Intelligence}}, Vol.~\bibinfo{volume}{37}.
  \bibinfo{pages}{13067--13075}.
\newblock
\href{https://doi.org/10.1609/aaai.v37i11.26535}{doi:\nolinkurl{10.1609/aaai.v37i11.26535}}


\bibitem[Liu et~al\mbox{.}(2024)]%
        {liu2024agentbench}
\bibfield{author}{\bibinfo{person}{Xiao Liu}, \bibinfo{person}{Hao Yu},
  \bibinfo{person}{Hanchen Zhang}, {et~al\mbox{.}}}
  \bibinfo{year}{2024}\natexlab{}.
\newblock \showarticletitle{{AgentBench}: Evaluating {LLM}s as Agents}. In
  \bibinfo{booktitle}{\emph{International Conference on Learning
  Representations}}.
\newblock


\bibitem[Mahmud and Kandogan(2026)]%
        {mahmud2026semanticgap}
\bibfield{author}{\bibinfo{person}{Jalal Mahmud} {and} \bibinfo{person}{Eser
  Kandogan}.} \bibinfo{year}{2026}\natexlab{}.
\newblock \showarticletitle{Exploring the Semantic Gap in Agentic Data Systems:
  A Formative Study of Operationalization Failures in Analytical Workflows}.
\newblock \bibinfo{journal}{\emph{arXiv preprint arXiv:2607.00828}}
  (\bibinfo{year}{2026}).
\newblock
\href{https://doi.org/10.48550/arXiv.2607.00828}{doi:\nolinkurl{10.48550/arXiv.2607.00828}}


\bibitem[Majumder et~al\mbox{.}(2025)]%
        {majumder2024discoverybench}
\bibfield{author}{\bibinfo{person}{Bodhisattwa~Prasad Majumder},
  \bibinfo{person}{Harshit Surana}, \bibinfo{person}{Dhruv Agarwal},
  \bibinfo{person}{Bhavana Dalvi~Mishra}, \bibinfo{person}{Abhijeetsingh
  Meena}, \bibinfo{person}{Aryan Prakhar}, \bibinfo{person}{Tirth Vora},
  \bibinfo{person}{Tushar Khot}, \bibinfo{person}{Ashish Sabharwal}, {and}
  \bibinfo{person}{Peter Clark}.} \bibinfo{year}{2025}\natexlab{}.
\newblock \showarticletitle{{DiscoveryBench}: Towards Data-Driven Discovery
  with Large Language Models}. In \bibinfo{booktitle}{\emph{International
  Conference on Learning Representations}}.
\newblock


\bibitem[Nahid et~al\mbox{.}(2026)]%
        {nahid2026rethinking}
\bibfield{author}{\bibinfo{person}{Md~Mahadi~Hasan Nahid},
  \bibinfo{person}{Davood Rafiei}, \bibinfo{person}{Weiwei Zhang}, {and}
  \bibinfo{person}{Yong Zhang}.} \bibinfo{year}{2026}\natexlab{}.
\newblock \showarticletitle{Rethinking Schema Linking: A Context-Aware
  Bidirectional Retrieval Approach for Text-to-{SQL}}. In
  \bibinfo{booktitle}{\emph{Findings of the Association for Computational
  Linguistics: {EACL} 2026}}. \bibinfo{publisher}{Association for Computational
  Linguistics}, \bibinfo{pages}{4516--4546}.
\newblock
\href{https://doi.org/10.18653/v1/2026.findings-eacl.236}{doi:\nolinkurl{10.18653/v1/2026.findings-eacl.236}}


\bibitem[{OpenAI}(2023)]%
        {openai2023gpt4}
\bibfield{author}{\bibinfo{person}{{OpenAI}}.} \bibinfo{year}{2023}\natexlab{}.
\newblock \showarticletitle{{GPT-4} Technical Report}.
\newblock \bibinfo{journal}{\emph{arXiv preprint arXiv:2303.08774}}
  (\bibinfo{year}{2023}).
\newblock
\href{https://doi.org/10.48550/arXiv.2303.08774}{doi:\nolinkurl{10.48550/arXiv.2303.08774}}


\bibitem[Robertson and Zaragoza(2009)]%
        {robertson2009bm25}
\bibfield{author}{\bibinfo{person}{Stephen Robertson} {and}
  \bibinfo{person}{Hugo Zaragoza}.} \bibinfo{year}{2009}\natexlab{}.
\newblock \showarticletitle{The Probabilistic Relevance Framework: {BM25} and
  Beyond}.
\newblock \bibinfo{journal}{\emph{Foundations and Trends in Information
  Retrieval}} \bibinfo{volume}{3}, \bibinfo{number}{4} (\bibinfo{year}{2009}),
  \bibinfo{pages}{333--389}.
\newblock
\href{https://doi.org/10.1561/1500000019}{doi:\nolinkurl{10.1561/1500000019}}


\bibitem[Rosenberg et~al\mbox{.}(2026)]%
        {rosenberg2026endoflife}
\bibfield{author}{\bibinfo{person}{Mara Rosenberg}, \bibinfo{person}{Irena
  Cenzer}, \bibinfo{person}{Alexander~K. Smith}, {and}
  \bibinfo{person}{Ashwin~A. Kotwal}.} \bibinfo{year}{2026}\natexlab{}.
\newblock \showarticletitle{End-of-Life Loneliness, Social Isolation, and
  Symptom Burden: A Nationally-Representative Study}.
\newblock \bibinfo{journal}{\emph{Journal of the American Geriatrics Society}}
  (\bibinfo{year}{2026}).
\newblock
\href{https://doi.org/10.1111/jgs.70398}{doi:\nolinkurl{10.1111/jgs.70398}}


\bibitem[Salton and Buckley(1988)]%
        {salton1988term}
\bibfield{author}{\bibinfo{person}{Gerard Salton} {and}
  \bibinfo{person}{Christopher Buckley}.} \bibinfo{year}{1988}\natexlab{}.
\newblock \showarticletitle{Term-Weighting Approaches in Automatic Text
  Retrieval}.
\newblock \bibinfo{journal}{\emph{Information Processing \& Management}}
  \bibinfo{volume}{24}, \bibinfo{number}{5} (\bibinfo{year}{1988}),
  \bibinfo{pages}{513--523}.
\newblock
\href{https://doi.org/10.1016/0306-4573(88)90021-0}{doi:\nolinkurl{10.1016/0306-4573(88)90021-0}}


\bibitem[Sonnega et~al\mbox{.}(2014)]%
        {sonnega2014hrs}
\bibfield{author}{\bibinfo{person}{Amanda Sonnega}, \bibinfo{person}{Jessica~D.
  Faul}, \bibinfo{person}{Mary~Beth Ofstedal}, \bibinfo{person}{Kenneth~M.
  Langa}, \bibinfo{person}{John W.~R. Phillips}, {and}
  \bibinfo{person}{David~R. Weir}.} \bibinfo{year}{2014}\natexlab{}.
\newblock \showarticletitle{Cohort Profile: the Health and Retirement Study
  ({HRS})}.
\newblock \bibinfo{journal}{\emph{International Journal of Epidemiology}}
  \bibinfo{volume}{43}, \bibinfo{number}{2} (\bibinfo{year}{2014}),
  \bibinfo{pages}{576--585}.
\newblock
\href{https://doi.org/10.1093/ije/dyu067}{doi:\nolinkurl{10.1093/ije/dyu067}}


\bibitem[Su et~al\mbox{.}(2025)]%
        {su2025aging}
\bibfield{author}{\bibinfo{person}{Chen Su}, \bibinfo{person}{Sen Zhang},
  \bibinfo{person}{Qiandan Zheng}, \bibinfo{person}{Jie Miao}, {and}
  \bibinfo{person}{Junhong Guo}.} \bibinfo{year}{2025}\natexlab{}.
\newblock \showarticletitle{Self-Perceptions of Aging and Sarcopenia in Older
  Adults: The Mediating Role of {IADL}}.
\newblock \bibinfo{journal}{\emph{Frontiers in Medicine}}  \bibinfo{volume}{12}
  (\bibinfo{year}{2025}), \bibinfo{pages}{1693158}.
\newblock
\href{https://doi.org/10.3389/fmed.2025.1693158}{doi:\nolinkurl{10.3389/fmed.2025.1693158}}


\bibitem[Taslimi et~al\mbox{.}(2025)]%
        {taslimi2025rq}
\bibfield{author}{\bibinfo{person}{Sina Taslimi}, \bibinfo{person}{Artemis
  Capari}, \bibinfo{person}{Hosein Azarbonyad}, \bibinfo{person}{Zi~Long Zhu},
  \bibinfo{person}{Zubair Afzal}, \bibinfo{person}{Evangelos Kanoulas}, {and}
  \bibinfo{person}{George Tsatsaronis}.} \bibinfo{year}{2025}\natexlab{}.
\newblock \showarticletitle{Extracting, Detecting, and Generating Research
  Questions for Scientific Articles}. In \bibinfo{booktitle}{\emph{Proceedings
  of the 31st International Conference on Computational Linguistics}}.
  \bibinfo{publisher}{Association for Computational Linguistics},
  \bibinfo{pages}{8573--8588}.
\newblock


\bibitem[Tsereteli et~al\mbox{.}(2022)]%
        {tsereteli2022svident}
\bibfield{author}{\bibinfo{person}{Tornike Tsereteli},
  \bibinfo{person}{Yavuz~Selim Kartal}, \bibinfo{person}{Simone~Paolo
  Ponzetto}, \bibinfo{person}{Andrea Zielinski}, \bibinfo{person}{Kai Eckert},
  {and} \bibinfo{person}{Philipp Mayr}.} \bibinfo{year}{2022}\natexlab{}.
\newblock \showarticletitle{Overview of the {SV-Ident} 2022 Shared Task on
  Survey Variable Identification in Social Science Publications}. In
  \bibinfo{booktitle}{\emph{Proceedings of the Third Workshop on Scholarly
  Document Processing}}.
\newblock
\href{https://doi.org/10.18653/v1/2022.sdp-1.29}{doi:\nolinkurl{10.18653/v1/2022.sdp-1.29}}


\bibitem[Tsereteli et~al\mbox{.}(2024)]%
        {tsereteli2024sil}
\bibfield{author}{\bibinfo{person}{Tornike Tsereteli}, \bibinfo{person}{Daniel
  Ruffinelli}, {and} \bibinfo{person}{Simone~Paolo Ponzetto}.}
  \bibinfo{year}{2024}\natexlab{}.
\newblock \showarticletitle{Enriching Social Science Research via Survey Item
  Linking}.
\newblock \bibinfo{journal}{\emph{arXiv preprint arXiv:2412.15831}}
  (\bibinfo{year}{2024}).
\newblock
\href{https://doi.org/10.48550/arXiv.2412.15831}{doi:\nolinkurl{10.48550/arXiv.2412.15831}}


\bibitem[Tu et~al\mbox{.}(2023)]%
        {tu2023sdrquerier}
\bibfield{author}{\bibinfo{person}{Yamei Tu}, \bibinfo{person}{Olga Li},
  \bibinfo{person}{Junpeng Wang}, \bibinfo{person}{Han-Wei Shen},
  \bibinfo{person}{Przemek Powa{\l}ko}, \bibinfo{person}{Irina Tomescu-Dubrow},
  \bibinfo{person}{Kazimierz~M. S{\l}omczy{\'n}ski}, \bibinfo{person}{Spyros
  Blanas}, {and} \bibinfo{person}{J.~Craig Jenkins}.}
  \bibinfo{year}{2023}\natexlab{}.
\newblock \showarticletitle{{SDRQuerier}: A Visual Querying Framework for
  Cross-National Survey Data Recycling}.
\newblock \bibinfo{journal}{\emph{IEEE Transactions on Visualization and
  Computer Graphics}} \bibinfo{volume}{29}, \bibinfo{number}{6}
  (\bibinfo{year}{2023}), \bibinfo{pages}{2862--2874}.
\newblock
\href{https://doi.org/10.1109/TVCG.2023.3261944}{doi:\nolinkurl{10.1109/TVCG.2023.3261944}}


\bibitem[Viswanathan et~al\mbox{.}(2023)]%
        {viswanathan2023datafinder}
\bibfield{author}{\bibinfo{person}{Vijay Viswanathan}, \bibinfo{person}{Kiril
  Gashteovski}, \bibinfo{person}{Carolin Lawrence}, \bibinfo{person}{Tongshuang
  Wu}, {and} \bibinfo{person}{Graham Neubig}.} \bibinfo{year}{2023}\natexlab{}.
\newblock \showarticletitle{{DataFinder}: Scientific Dataset Recommendation
  from Natural Language Descriptions}. In \bibinfo{booktitle}{\emph{Proceedings
  of ACL}}. \bibinfo{pages}{10288--10303}.
\newblock
\href{https://doi.org/10.18653/v1/2023.acl-long.573}{doi:\nolinkurl{10.18653/v1/2023.acl-long.573}}


\bibitem[Wang et~al\mbox{.}(2020)]%
        {wang2020ratsql}
\bibfield{author}{\bibinfo{person}{Bailin Wang}, \bibinfo{person}{Richard
  Shin}, \bibinfo{person}{Xiaodong Liu}, \bibinfo{person}{Oleksandr Polozov},
  {and} \bibinfo{person}{Matthew Richardson}.} \bibinfo{year}{2020}\natexlab{}.
\newblock \showarticletitle{{RAT-SQL}: Relation-Aware Schema Encoding and
  Linking for Text-to-{SQL} Parsers}. In \bibinfo{booktitle}{\emph{Proceedings
  of the 58th Annual Meeting of the Association for Computational
  Linguistics}}. \bibinfo{pages}{6778--6788}.
\newblock
\href{https://doi.org/10.18653/v1/2020.acl-main.677}{doi:\nolinkurl{10.18653/v1/2020.acl-main.677}}


\bibitem[Wang et~al\mbox{.}(2025)]%
        {wang2025linkalign}
\bibfield{author}{\bibinfo{person}{Yihan Wang}, \bibinfo{person}{Peiyu Liu},
  {and} \bibinfo{person}{Xin Yang}.} \bibinfo{year}{2025}\natexlab{}.
\newblock \showarticletitle{{LinkAlign}: Scalable Schema Linking for Real-World
  Large-Scale Multi-Database Text-to-{SQL}}. In
  \bibinfo{booktitle}{\emph{Proceedings of the 2025 Conference on Empirical
  Methods in Natural Language Processing}}. \bibinfo{pages}{977--991}.
\newblock
\href{https://doi.org/10.18653/v1/2025.emnlp-main.51}{doi:\nolinkurl{10.18653/v1/2025.emnlp-main.51}}


\bibitem[Wang et~al\mbox{.}(2026)]%
        {wang2026autolink}
\bibfield{author}{\bibinfo{person}{Ziyang Wang}, \bibinfo{person}{Yuanlei
  Zheng}, \bibinfo{person}{Zhenbiao Cao}, \bibinfo{person}{Xiaojin Zhang},
  \bibinfo{person}{Zhongyu Wei}, \bibinfo{person}{Pei Fu},
  \bibinfo{person}{Zhenbo Luo}, \bibinfo{person}{Wei Chen}, {and}
  \bibinfo{person}{Xiang Bai}.} \bibinfo{year}{2026}\natexlab{}.
\newblock \showarticletitle{{AutoLink}: Autonomous Schema Exploration and
  Expansion for Scalable Schema Linking in Text-to-{SQL} at Scale}. In
  \bibinfo{booktitle}{\emph{Proceedings of the AAAI Conference on Artificial
  Intelligence}}, Vol.~\bibinfo{volume}{40}. \bibinfo{pages}{33809--33817}.
\newblock
\href{https://doi.org/10.1609/aaai.v40i40.40672}{doi:\nolinkurl{10.1609/aaai.v40i40.40672}}


\bibitem[Xiao et~al\mbox{.}(2024)]%
        {xiao2023cpack}
\bibfield{author}{\bibinfo{person}{Shitao Xiao}, \bibinfo{person}{Zheng Liu},
  \bibinfo{person}{Peitian Zhang}, \bibinfo{person}{Niklas Muennighoff},
  \bibinfo{person}{Defu Lian}, {and} \bibinfo{person}{Jian-Yun Nie}.}
  \bibinfo{year}{2024}\natexlab{}.
\newblock \showarticletitle{{C-Pack}: Packed Resources for General Chinese
  Embeddings}. In \bibinfo{booktitle}{\emph{Proceedings of the 47th
  International ACM SIGIR Conference on Research and Development in Information
  Retrieval}}. \bibinfo{publisher}{Association for Computing Machinery},
  \bibinfo{pages}{641--649}.
\newblock
\href{https://doi.org/10.1145/3626772.3657878}{doi:\nolinkurl{10.1145/3626772.3657878}}


\bibitem[Yang et~al\mbox{.}(2024)]%
        {yang2024llmmeasure}
\bibfield{author}{\bibinfo{person}{Yi Yang}, \bibinfo{person}{Hanyu Duan},
  \bibinfo{person}{Jiaxin Liu}, {and} \bibinfo{person}{Kar~Yan Tam}.}
  \bibinfo{year}{2024}\natexlab{}.
\newblock \showarticletitle{{LLM-Measure}: Generating Valid, Consistent, and
  Reproducible Text-Based Measures for Social Science Research}.
\newblock \bibinfo{journal}{\emph{arXiv preprint arXiv:2409.12722}}
  (\bibinfo{year}{2024}).
\newblock
\href{https://doi.org/10.48550/arXiv.2409.12722}{doi:\nolinkurl{10.48550/arXiv.2409.12722}}


\end{thebibliography}
\end{document}